# Computational accelerator science needs towards laser-plasma accelerators for future colliders

C.G.R. Geddes, J.-L. Vay, C.B. Schroeder, E. Esarey, W. P. Leemans (LBNL)

Laser plasma accelerators (LPAs)[1,2] have the potential to reduce the size of future linacs for high energy physics by more than an order of magnitude, due to their high gradient. Research is in progress at current facilities, including the BELLA PetaWatt laser at LBNL, towards high quality 10 GeV beams and staging of multiple modules, as well as control of injection and beam quality. The path towards high-energy physics applications will likely involve hundreds of such stages, with beam transport, conditioning and focusing [3]. Current research focuses on addressing physics and R&D challenges required for a detailed conceptual design of a future collider, as detailed in the Snowmass 2013 white paper by J.P. Delahaye et al. Here, the tools used to model these accelerators and their resource requirements are summarized, both for current work and to support R&D addressing issues related to collider concepts.

Simulations of LPAs [4] must resolve dynamics of plasma ionization and formation, laser propagation and energy transfer in meter-scale plasmas, and the injection and evolution of high quality particle beams in the plasma and laser fields. The core methods for current accelerator simulations are explicit and implicit particle in cell and fluid models. These codes successfully model current experiments, at the 1-10 GeV level in m-scale plasmas, with 0.1 micron level emittance, and including loading of the plasma by the accelerating beam. The separation of spatial and time scales between the micron laser period and the centimeter to meter scale acceleration distance, and the requirement to resolve high quality beams, means these simulations stretch current computational capabilities.

Parallelization is typically via domain decomposition which, when using the most common second-order finite difference time domain methods on staggered grids, scale well to > 100 kcores. This is limited by area-to-volume ratio. Envelope codes in some cases also require parallel matrix solves. An important limit is I/O which typically does not scale well and must be addressed. Solutions include improved bandwidth and I/O tools that allow in-situ diagnostics. Development of new tools is an active research area, and recently includes computation in a boosted frame, where the scale disparity is reduced, envelope codes which average over the laser period, and r-z codes which can in certain cases be used in place of full 3D codes. These mitigate cost and have allowed modeling at the 10 GeV level [4,5].

Accelerator simulations take input from MHD plasma formation models (presently 1D) [6], which in turn take input from gas dynamic simulations to determine the initial state. This sequence is enabled by the separation of timescales between the relevant phenomena. Radiation by the particles is modeled with separate codes that implement particle tracking and interpolation of trajectories in order to resolve short-wavelength radiation [7].

New capabilities will be required to support development of plasma accelerators to address high-energy physics applications [3]. While research on 10 GeV stages with beam loading can use existing capabilities, other R&D requires new tools:
- staging of order 100 LPAs, each of meter length,
- resolution of 10 nm-scale emittances, in injection, cooling, and acceleration,
- scattering and radiation effects, both in the plasma and between stages,
- positron production and acceleration in intense plasma fields,
- novel beam cooling methods including plasma based radiation generation,
- survival of spin polarization,
- plasma based final focus (e.g. adiabatic plasma lens) to reduce focal length.

These needs simultaneously increase the length of simulation and the accuracy with which beam emittance must be resolved by one to two orders of magnitude, while domain size increases only modestly. They also require adding physical models (such as positron production, polarization, radiation), which exist separately but will need to be computed self consistently with plasma and beam evolution. To correctly address contributions to emittance, radiation modeling must resolve quantization. The number of simulated particles will also have to be significantly increased to correctly capture statistics of radiation, scattering, and polarization.

Research on tailoring of the plasma to control injection or dephasing, and on near-hollow channels to mitigate scattering-induced emittance growth require that current 1D plasma formation codes be developed to account for 3D effects. For some concepts, laser heat deposition should be modeled self consistently instead of serially. The requirements of high average power at kHz-MHz repetition rates will also be important, including:
- heat flow in the target and its effects on plasma formation,
- modeling to support high average power laser development.

While weak scaling can increase particle number and resolution, needs for increased run length and accuracy with added physical models motivate new methods, especially on emerging architectures. Emerging multicore or SIMD systems typically function well with PIC codes, subject to memory needs, and selected implementations now exist. At this point, each system requires specialized development. Hence, high priorities to make such systems productive would include having sufficient notice of the chosen architecture with a test bed available, and development of common compilers and tools. Heterogeneous decomposition will also likely be required. Development of common libraries for I/O that are scalable and can exploit the full bandwidth of the system are needed. Parallel analysis tools incorporating advanced mathematics are also important. Lastly, numerical methods with improved accuracy and reduced unphysical momentum contributions [8] will be critical. These may include Vlasov or other models. In particular, as compute power appears likely to increase faster than bandwidth, more accurate methods allowing longer timesteps (even if at higher computational cost) may be advantageous.